\documentclass[12pt]{iopart}

\usepackage{iopams}

\usepackage{graphicx,epsf,bm}
\usepackage{color}
\usepackage{epstopdf}
\usepackage{soul}

\newcommand{\red}{\textcolor{black}}

\begin{document}
\title{Asymmetric Landau bands due to  spin-orbit coupling}

\author{Sigurdur I. Erlingsson$^1$, Andrei Manolescu$^1$, and D. C. Marinescu$^2$}
\address{$^1$ School of Science and Engineering, Reykjavik University, Menntavegur 1, 
IS-101 Reykjavik, Iceland}
\address{$^2$ Department of Physics and Astronomy, Clemson University,Clemson, South Carolina 29634, USA}


\begin{abstract}
We show that the Landau bands obtained in a two-dimensional lateral
semiconductor superlattice with spin-orbit coupling (SOC) of the
Rashba/Dresselhaus type, linear in the electron momentum, placed in
a tilted magnetic field, do not follow the symmetry of the spatial
modulation. Moreover, this phenomenology is found to depend on the
relative tilt of magnetic field and on the SOC type: a) when
only Rashba SOC exists and the magnetic field is tilted in the direction
of the superlattice b) Dresselhaus SOC exists and the magnetic field is
tilted in the direction perpendicular to the superlattice. Consequently,
measurable properties of the modulated system become anisotropic in a
tilted magnetic field when the field is conically rotated around the
$z$ axis, at a fixed polar angle, as we demonstrate by calculating the
resistivity and the magnetization.
\end{abstract}

\maketitle

\section{Introduction}

For more than ten years, the complex phenomenology observed in two
dimensional semiconductor structures with spin-orbit coupling (SOC)
has been at the forefront of numerous experimental and theoretical
explorations \cite{zutic04}. Originating in the asymmetry of the quantum
well or in the inversion asymmetry of the crystal structure, the spin
orbit interaction, of the Rashba \cite{rashba} and Dresselhaus
\cite{dresselhaus} type, respectively, describes the coupling
of the electron spin to its momentum, generating the possibility
of simultaneous spin and charge effects, as well as the possibility
of controlling the spin through the application of electric fields
\cite{Engels1997,Nitta1997}.  It is the latter reason that underlies
the continuous interest, because of a potential realization of semiconductor
based spintronic devices.  From this perspective, the investigation
of magnetotransport properties was focused from the very beginning on
identifying the best set-up which, in the presence of electric or magnetic
fields, can showcase spin-dependent macroscopic response functions,
and in particular spin or spin-polarized currents.

The transport theory of Landau levels in the presence of
electrostatic modulation goes already back a quarter of a century
\cite{winkler89,Zhang1990}.  Since then these systems have attracted continued
attention. In the more recent years of spintronics the effects of SOC were added
\cite{wang05}, modulation effects on the spin Hall conductance were
studied \cite{islam13}, and the effects of a tilted magnetic
field were considered \cite{dosSantos13}.  As has been pointed out,
the spin-orbit effect can be relatively enhanced in a tilted magnetic field,
as observed both in transport \cite{desrat05} and magnetization
measurements \cite{wilde13,Rupprecht2013}.  The origin of this enhancement
is that at Landau level crossing, where the orbital and Zeeman energies
are equal, the Landau levels are linear with the SOC strength, and the
spin mixing is maximized \cite{wilde13}.  The suppression of orbital
energy can also be achieved in one dimensional systems with periodic
modulation where it has been proposed that signature of SOC can be
measured in two terminal transport \cite{thorgilsson12}.

The experimental detection of the SOC in a nanostructure is in general
difficult.  The response functions depend on the energy dispersion,
which usually include some weak effects of the SOC.  An interesting
situation was described in Ref. \cite{Fedorov2005} where the energy
spectrum of a two-dimensional (2D), laterally symmetric quantum wire, was
calculated. Due to the interplay of SOC and an in-plane magnetic field
perpendicular to the length of a wire the energy dispersion becomes
asymmetric with respect to the sign reversal of the momentum along
the wire \cite{Pershin2004}.  The authors of Ref. \cite{Fedorov2005} predicted that this
situation can be experimentally verified by phototransport measurements,
where an asymmetric band structure is essential for obtaining photocurrents
\cite {Belkov2005}.  Nevertheless, to our knowledge such experiments are not 
reported for quantum wires or another nanostructure with lateral
confinement, where the electrons are the current carriers. Photoconductivity
had been studied in the two-dimensional electron gas with SOC,
but based on a different mechanism, of an imbalanced population of
electrons with different spin orientation under the action of light
and an in-plane magnetic field \cite{Ganichev2003}.  Only in a recent
study magnetophotocurrents with possible spin polarization have been
generated by electron-hole excitations in asymmetric Rashba spin-split
bands in layered semiconductor BiTeI \cite{Ogawa2013}.
More complicated asymmetries of the energy spectra, for example due to the simultaneous
presence of Rashba and Dresselhaus SOC in thick quantum wells in magnetic field,
can generate equilibrium spin currents which are still not observed in 
experiments \cite{Nakhmedov2011}.

In the present paper we propose alternative possibilities to detect the
effect on the energy dispersion resulting from the combination of SOC
and an in-plane magnetic field in a 2D electron system (2DES).  We consider a
tilted magnetic field with a vertical component that is strong enough to
induce the quantization of the Landau levels in the electron plane. We
also consider a lateral modulation in the plane obtained by applying a
periodic electrostatic potential along one direction in the plane,
as shown in Fig. \ref{fig:System}.  We show that in the presence of
SOC the resulting Landau bands, and implicitly the density of states,
depend on the orientation of the in-plane magnetic field relatively
to the direction of the modulation.  In particular, if the electrostatic
modulation has an inversion symmetry, the Landau bands do not obey
it.  We calculate the conductivity tensor and the total energy of the
modulated 2DES and show that the presence of the SOC could be observed
in magnetotransport and magnetization experiments, in a tilted magnetic
field rotating conically around the normal direction to the plane.

\section{The weakly modulated 2DES in a tilted field}

\begin{figure}
\begin{center}
\includegraphics[width=0.48\textwidth,angle=0]{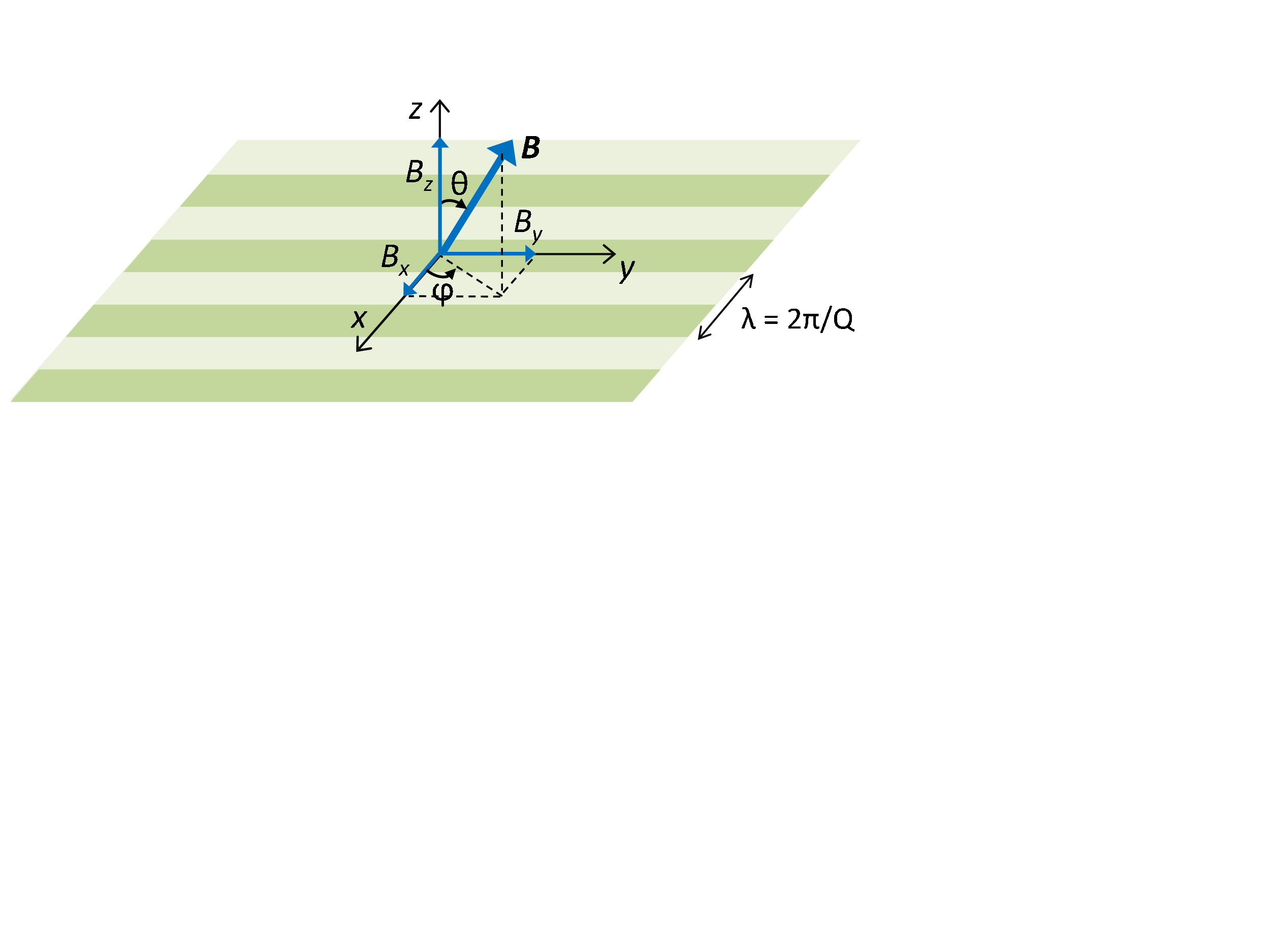}
\end{center}
\caption{(Color online) A 2DES situated in the $x-y$ plane exhibits a
density modulation (stripes) when subjected to a periodic potential $V_0\cos Qx$.
The magnetic field, represented by the blue thick arrows, is oriented along a 
direction determined by the polar and azimuthal angles, $\theta$
and $\varphi$, respectively.}
\vspace{5mm}
\label{fig:System}
\end{figure}

In the following discussion we consider the 2DES
situated in the $\hat{x}-\hat{y}$ plane, and subjected to a lateral
modulation along the $\hat{x}$ direction which is induced by applying a periodic
potential, $V(x)=V_0 \cos Qx$. A magnetic field of magnitude $B$, tilted
in respect with the $\hat{z}$ axis is also applied. Its direction is
described by the usual angular representation, $\bm{B}=B(\cos\varphi
\sin \theta,\sin\varphi \sin \theta,\cos \theta)$, as shown in Fig.\
\ref{fig:System}. We assume that the perpendicular component of the
magnetic field is strong enough such that the quantization of the single
particle energies into Landau levels is induced. With $a$ and $a^\dag$
the usual destruction and creation operators of the harmonic oscillator,
the single particle Hamiltonian is written in a spin basis provided by
the Pauli matrices, $\{\sigma_x,\sigma_y,\sigma_z\}$, as,
\begin{equation}
H_\mathrm{2D}=\hbar \omega_c (a^\dagger a+1/2)+\frac{g\mu B}{2}\bm{B}\cdot \bm{\sigma}
+V(x+\xi),
\label{eq:Hper}
\end{equation}
where $\xi=-\ell_c^2 p_y/\hbar$ is the so called center coordinate, 
$\ell_c=\sqrt{\hbar/eB}$ is the magnetic length, 
and a unitary transformation 
$e^{i p_x \xi/\hbar}$ was used to displace the origin of the modulation 
potential. In the absence of SOC the orbital and spin states decouple, the spin
states being either parallel or anti-parallel with the total magnetic field,
 \begin{eqnarray}
|\uparrow_{\bm{B}} \rangle&=& \cos\frac{\theta}{2}e^{i\frac{\varphi}{2}}|\uparrow \rangle
+\sin \frac{\theta}{2} e^{-i\frac{\varphi}{2}}|\downarrow \rangle \ , \nonumber\\
|\downarrow_{\bm{B}} \rangle&=& -\sin\frac{\theta}{2}e^{i\frac{\varphi}{2}}|\uparrow \rangle
+\cos \frac{\theta}{2} e^{i\frac{\varphi}{2}}|\downarrow \rangle \ ,
\label{Eigenvectors}
 \end{eqnarray}
where $|\uparrow \rangle$ and $|\downarrow \rangle$ are the usual
eigenstates of $\sigma_z$.  For a modulation amplitude much smaller than
the cyclotron energy, i. e. $V_0 \ll \hbar\omega_c$, the orbital states are
obtained in first order perturbation theory to be,
\begin{eqnarray}
|n,\xi\rangle &\approx& |n \rangle-\frac{V_0 \sin Q \xi }{\hbar \omega_c} 
(S_{n,n-1} |n-1\rangle- S_{n,n+1} |n+1\rangle ) \nonumber \\
&-&\frac{V_0 \cos Q \xi }{2\hbar \omega_c} C_{n,n-2} |n-2\rangle-C_{n,n+2} |n+2 \rangle),
\label{eq:Lbands}
\end{eqnarray}
where $S_{n,m}=\langle n|\sin Qx|m \rangle$ and $C_{n,m }=\langle n|\cos
Qx|m \rangle$ are matrix elements in the basis of the usual harmonics
oscillator states $|n \rangle$.  By increasing the tilt of the magnetic
field, such that the relative contribution of the in-plane component to the Zeeman
splitting is enhanced, opposite-spin Landau bands cross \cite{gfg85prb}.

In the presence of the Rashba interaction, of strength $\alpha$, 
the single-particle Hamiltonian acquires an additional term which reflects the coupling between the spin $\sigma$ and the electron momentum written in terms of the harmonic oscillator operators,
\begin{eqnarray}
H_\mathrm{R}&=&
\frac{\alpha}{\sqrt{2}\ell} (a \sigma_+ + a^\dagger \sigma_-) \ ,
\label{eq:HR}
\end{eqnarray}
where $\sigma_{\pm}=\sigma_x \pm i\sigma_y$ .
In these conditions the particle spin is not a good quantum number
and the energy bands couple in the spin space.  They can be determined
by projecting the full Hamiltonian $H_\mathrm{2D}+H_\mathrm{R}$ onto
the subspace $|n,\xi \rangle |\uparrow_{\bm{B}} \rangle$ and $|n-1,\xi
\rangle |\downarrow_{\bm{B}} \rangle$, followed by the diagonalization of
the resulting $2\times 2$ matrix.  $H_{2D}$ is trivially diagonal,
with diagonal elements
\begin{eqnarray}
\varepsilon^0_{n,\uparrow_{\bm{B}} }&=& \hbar \omega_c(n+1/2)-\frac{g\mu B}{2}+V_n\cos Q \xi \nonumber \\
\varepsilon^0_{n-1,\downarrow_{\bm{B}} }&=& \hbar \omega_c(n-1/2)+\frac{g\mu B}{2}+V_{n-1}\cos Q \xi
\label{Lbands0}
\end{eqnarray}
which are the standard Landau bands in the first order of the modulation strength, 
where $V_n=V_0 C_{n,n}$.

The Rashba term generates
\begin{eqnarray}
\frac{H_{R,n}}{\hbar \omega_c}&=&\frac{\alpha n\cos^2\frac{\theta}{2}}{\sqrt{2}\ell_c} 
(e^{i\varphi}\sigma_+ + \mbox{h.c.\ } )
+\frac{\alpha }{\sqrt{2}\ell_c}\frac{V_0}{\hbar \omega_c} \frac{\mathcal{S}_{n}+\mathcal{S}_{n-1}}{2}\nonumber \\
&+& \frac{\alpha }{\sqrt{2}\ell_c}\frac{V_0}{\hbar \omega_c} (\mathcal{O}_{z,n} \sigma_z+\mathcal{O}_{-,n} \sigma_+ +\mathcal{O}^*_{-,n}\sigma_- )\ ,
\label{eq:HRn}
\end{eqnarray}
where $\mathcal{S}_{n}=\sqrt{n}S_{n,n-1}-\sqrt{n+1}S_{n,n+1}$ and
\begin{equation}
\mathcal{O}_{z,n}=-\frac{\mathcal{S}_{n}-\mathcal{S}_{n-1}}{2}\sin\theta \cos \varphi \sin Q\xi\ ,
\label{eq:Ez}
\end{equation}
\begin{equation}
\mathcal{O}_{-,n}=\frac{\sqrt{n-1}C_{n,n-2}}{2}\sin^2\frac{\theta}{2} e^{-i\varphi}\cos Q \xi\ .
\label{eq:Em}
\end{equation}
In the absence of the periodic potential ($V_0=0$) 
the Rashba contribution reduces to the first term of Eq.\ (\ref{eq:HRn}) which, for a 
perpendicular  magnetic field ($\theta=0$) leads to the 
known exact eigenstates, since the Rashba interaction only couples two adjacent Landau
levels \cite{Winkler}.  As long as we are working in the regime $\hbar
\omega_c \gg V_0 \gg \frac{\alpha}{\sqrt{2}\ell}$ the Landau bands due
the effective modulation are obtained by diagonalizing each $2\times 2$
subspace for Landau level $n$ \cite{Erlingsson2010}.
\red{The other terms of Eq.\ (\ref{eq:HRn}) are generated by the periodic potential.  
The second term is simply an energy shift which depends on the Landau level 
(via $\mathcal{S}_n$).  The terms containing $\mathcal{O}_{z,n}$ and
$\mathcal{O}_{-,n}$ can be interpreted as, respectively, the perpendicular
and in-plane components of an effective, Landau level dependent magnetic
modulation that appears due to the interplay of the Rashba SOI and the
modulation potential.  It is clear from Eq. (7) that the perpendicular
component of the effective magnetic modulation leads to Landau bands that are
asymmetric around $Q\xi = 0$ or $Q\xi = \pi$, on account of the $\sin Q
\xi $ term, although the modulation potential is an even function of $x$.}
Symmetric bands, i. e. even functions of $\xi$, 
are recovered only in the case where $\cos\varphi=0$, i. e.\ for $\varphi=\pi/2$. 

In the case of Dresselhaus type of SOC, of strength $\beta$, which can
be written in a form equivalent to the Rashba coupling by using the
spin rotation $e^{i \frac{\pi}{2}\sigma_x}e^{i \frac{\pi}{4}\sigma_z}$,
one obtains very similar results for the Landau bands.  Due to this well
know Rashba-Dresselhaus transformation, the Landau bands obtained with
$\alpha$, $\beta$, and $(B_x,B_y,B_z)$ are identical to those obtained
with $\beta$, $\alpha$, and $(B_y,B_x,-B_z)$.

To conclude this section we notice that, according to
Eqs. (\ref{eq:Ez})-(\ref{eq:Em}), the amplitude of the Landau bands
depends on the strength of the SOC and on the orientation of the in-plane
magnetic field relatively to the modulation direction.  In particular,
if the external potential is an even function of $x$, the expected even
parity of the Landau bands in the Brillouin zone is broken by the Rashba
SOC if $B_x\neq 0$, or by the Dresselhaus SOC if if $B_y\neq 0$.

\section{The strongly modulated 2DES}

In order to increase the effect of the SOC on the energy dispersion we
consider a modulation potential with an amplitude much larger that the
cyclotron energy which can generate higher harmonics in the dispersion
of the energy bands.  To be more realistic we also include in the band
structure calculations the Coulomb interaction between the electrons. We
use the Hartree-Fock approximation where the Fock exchange is screened with
the static polarizability.  The long-range direct Coulomb term tends to
screen the external potential and reduce the energy dispersion of the
Landau bands, whereas the short-range exchange term tends to increase
the dispersion by contributing with negative energy to the occupied
states \cite{Manolescu1997}.  

\begin{figure}
\begin{center}
\includegraphics[width=0.75\textwidth,angle=0]{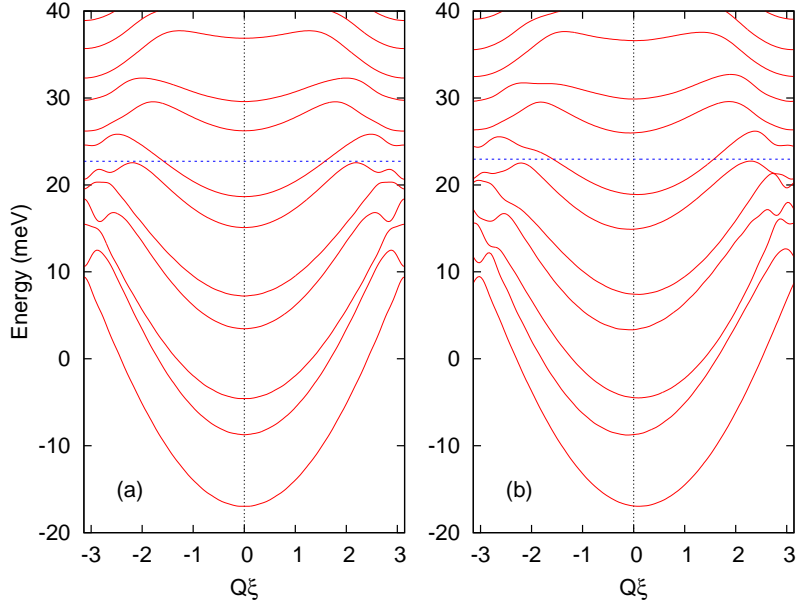}
\end{center}
\vspace{-5mm}
\caption{(Color online) Landau bands created by a modulation potential
along the $x$ direction, with period $\lambda=150$ nm and amplitude
$V_0=40$ meV, vs. the center coordinate $\xi$. The total magnetic field
is $B=9.02$ T and the polar angle $\theta=80.2^o$.  (a) The magnetic
field is tilted in the $y$ direction, i. e. $B_x=0$, or $\varphi=\pi/2$, and
the Landau bands are symmetric within the Brillouin zone $-\pi < Q\xi <
\pi$, where $Q=2\pi/\lambda$.  (b) The magnetic field is tilted in the
$x$ direction, i. e. $B_y=0$, or $\varphi=0$, and the Landau bands become
asymmetric. This asymmetry is due to the Rashba SOC. The horizontal dashed
line indicates the Fermi energy. The material parameters correspond to InAs,
with SOC $\alpha=20$ meV nm and $\beta=0$}
\vspace{5mm}
\label{fig:Landau_bands}
\end{figure}

Examples of resulting Landau bands are shown in
Fig. \ref{fig:Landau_bands}, for a modulation potential of period
$\lambda=150$ nm and amplitude $V_0=40$ meV.  The material parameters are
chosen as for InAs: effective mass $m_{\rm eff}=0.023 m_0$, effective g-factor
$g_{\rm eff}=-14.9$, Rashba SOC strength $\alpha=20$ meV nm. 
The density of electrons is $2.4 \times 10^{11} \ {\rm cm^{-2}}$.  In order to
maximize the effect of the SOC we tilt the magnetic field such that the
Zeeman and cyclotron energies are equal in the absence of the modulation,
which allows the strongest mixing of spin up and spin down states.
The corresponding polar angle is $\theta=80.2^o$.  The total magnetic
field is $B=9.02$ T.  The Brillouin zone corresponds to center coordinate
interval $(-\pi/Q, \pi/Q)$.  As expected from the perturbative limit,
if the magnetic field is tilted in the $y$ direction only, the bands
are symmetric relatively to the center of the Brillouin zone, i. e. they
are even functions of $\xi$.  Whereas if the magnetic field is tilted along $x$
the bands become asymmetric, due to the Rashba SOC.

Consequently observables that depend on the energy structure should be
anisotropic relatively to a rotation of the magnetic field around the
$z$ axis, i. e. with a fixed polar angle $\theta$.  In the following
we describe such results for the electrical conductivity and for the
magnetization of the modulated 2DES endowed with SOC.

\section{Conductivity}

We performed the calculations of the conductivity tensor,
$\sigma_{xx}, \sigma_{yy}, \sigma_{xy}$, using the Kubo linear response formula.
The longitudinal conductivity along the modulation, $\sigma_{xx}$, depends on the
collision-induced broadening of the Landau bands which is taken into account within the
self-consistent Born approximation, assuming delta-scatterers \cite{Zhang1990}.
Following the result for the uniform 2DES we assume a broadening
$\Gamma=\gamma\sqrt{B_z}$, with $\gamma=0.15$ meV/T$^{1/2}$.  The longitudinal
conductivity parallel to the modulation strips, $\sigma_{yy}$, is dominated by
the drift of the electrons along the strips, which is known as band conductivity,
but also incorporates a weaker scattering component \cite{Zhang1990}.
The scattering conductivity is (approximately) proportional to the density of states 
at the Fermi level squared and therefore its maxima correspond to the van Hove 
singularities of the energy bands.  In contrast, the band conductivity
is inversely proportional to the same quantity, and therefore has maxima in the 
middle of the Landau bands. \red{In a similar manner the scattering conductivity is
proportional to $\Gamma^2$ whereas the band conductivity is proportional to 
$\Gamma^{-2}$} \cite{Manolescu1997}.

\begin{figure}
\begin{center}
\includegraphics[width=1.00\textwidth,angle=0]{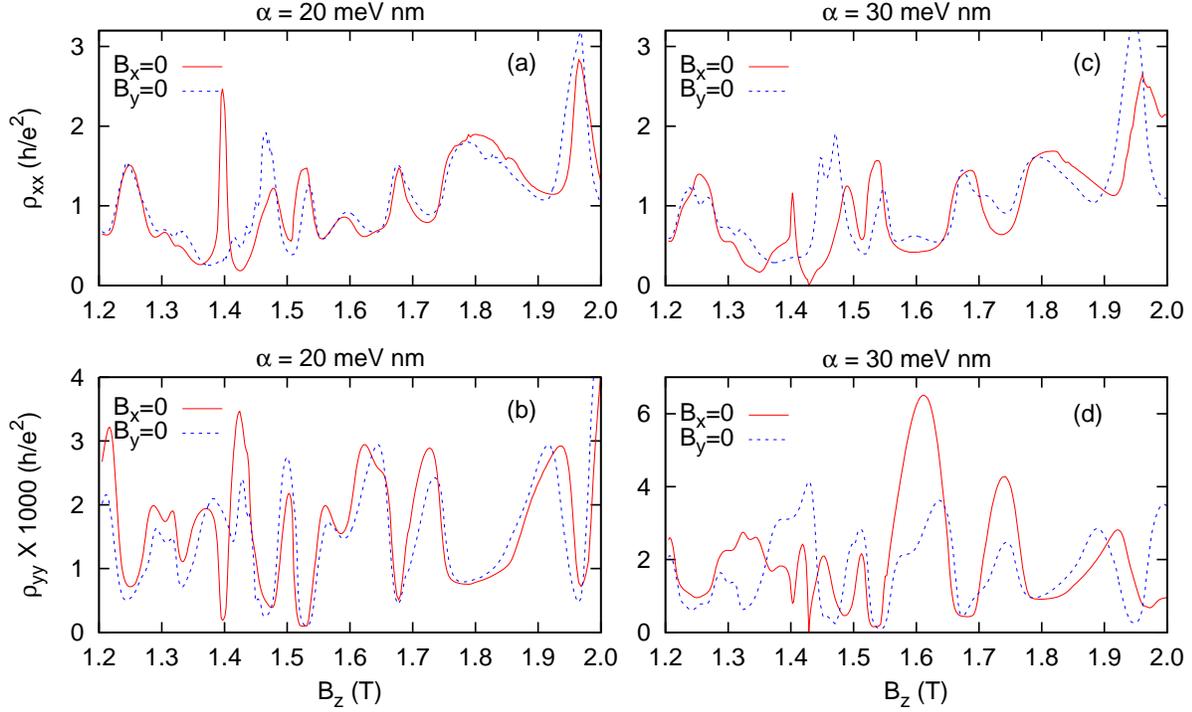}
\end{center}
\vspace{-8mm}
\caption{(Color online) Resistivities of the modulated 2DES compared, by
tilting the magnetic field along the $y$ or $x$ direction, i. e. when
$\varphi=\pi/2$ ($B_x=0$) and $\varphi=0$ ($B_y=0$), respectively.  (a) The
resistivity along the modulation, i. e. in the $x$ direction. (b)  The
resistivity in the homogeneous direction $y$, magnified 1000 times.
(c) and (d) The resistivities recalculated with the Rashba SOC parameter increased by 50\%.
In both cases the Dresselhaus SOC strength is $\beta=3$ meV nm. 
}
\vspace{8mm}
\label{fig:Resistivity}
\end{figure}

We treat the Hall conductivity in the simplest possible way, with
a filling-factor ($\nu$) dependent formula, $\sigma_{xy}=(e^2/h) \nu$, which does not describe
Hall plateaus, but it is still a qualitatively reasonable result for a strong modulation.
Since usually in transport experiments the resistivities are measured, rather than
conductivities, we display in Fig.\ \ref{fig:Resistivity} the longitudinal resistivities
\begin{equation}
\rho_{xx,yy}=\frac{\sigma_{yy,xx}}{\sigma_{xx}\sigma_{yy}+\sigma_{xy}^2}\ .
\end{equation}
In general $\sigma_{xx}\sigma_{yy} \ll \sigma_{xy}^2$
\cite{Zhang1990,Manolescu1997} and therefore $\rho_{xx}\sim \sigma_{yy}$
and $\rho_{yy}\sim \sigma_{xx}$.  \red{On account of the opposite relation
between conductivities and density of states the two resistivities
oscillate with the magnetic field in antiphase, as can be seen by comparing 
Figs. \ref{fig:Resistivity} a-b or c-d, respectively. For example,} 
in Fig. \ref{fig:Landau_bands}
the $z$ component of the magnetic field is $B_z=B\cos\theta=1.54$ T.
Correspondingly, in Fig. \ref{fig:Resistivity} we notice a
succession of a local maximum and a local minimum for $1.50 {\rm T} <
B_z < 1.55 {\rm T}$ for $\rho_{xx}$ (which is dominated by the band
conductivity via $\sigma_{yy}$), since the density of states at the
Fermi level increases by approaching the top of a Landau band, and
the other order, first a minimum and then a maximum for $\rho_{yy}$
(which is dominated by the scattering conductivity $\sigma_{xx}$).
\red{The large dispersion of the Landau bands and the relatively small
disorder broadening determine a value for $\rho_{yy}$ that is three
orders of magnitude smaller than $\rho_{xx}$.  }

\red{In the transport calculations we included a Dresselhaus SOC
with $\beta=3$ meV nm, but that has no significant contribution in
Fig. \ref{fig:Resistivity}.  More important, by increasing the Rashba
SOC strength, from $\alpha=20$ to $\alpha=30$ meV nm, the resistivities
change quite a lot. Both dispersion and asymmetry of the Landau bands increase with 
increasing SOC.} The
behavior of the resistivities is complex and rather difficult
to interpret.  Nevertheless, as expected from the Landau band structure,
the message is that both longitudinal resistivities ($xx$ and $yy$)
should depend on the orientation of the magnetic field projection in
the plane of the 2DES, i. e. on the azimuthal angle $\varphi$.

\section{Magnetization}

The magnetization ${\bf M}$ of the modulated 2DES is
related to the total free energy of the electrons, $E-TS$, by the thermodynamic relation
\begin{equation}
d(E-TS)=-{\bf M}d{\bf B} \,
\label{eq:MdB}
\end{equation}
where $E$ is the total energy and $S$ the entropy of the 2DES.  In the
zero-temperature limit we can evaluate the measurable magnetization in
the direction perpendicular to the 2DES in a thermodynamic manner \cite{Gudmundsson2000}: 
\begin{equation}
M_z=-\frac{\partial E}{\partial B_z} \ .
\end{equation}
The total energy was calculated from the Landau bands as the sum -- over all
occupied effective single-particle states -- of the single-particle energies
obtained in the absence of the Coulomb interaction, plus one half of
the Coulomb energy in order to exclude the double counting of electrons, 
\red{like in Eq.\ 3 of Ref. \cite{Meinel2001}. Then, according to Eq.\ \ref{eq:MdB}, the magnetization 
was obtained with a numerical derivative procedure, on a sufficiently dense grid
of magnetic field values.}

\begin{figure}
\begin{center}
\includegraphics[width=0.85\textwidth,angle=0]{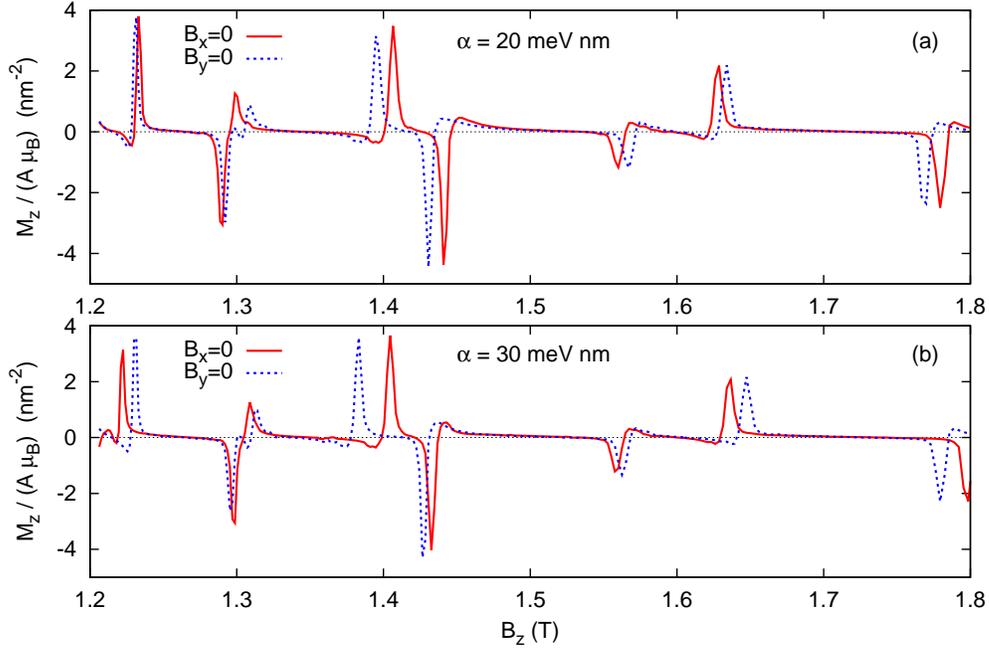}
\end{center}
\vspace{-8mm}
\caption{(Color online) (a) The magnetization in the $z$ direction, in units
of Bohr magneton $\mu_B$, and per surface area $A$ in 
nm$^{2}$, vs. the $z$ component of the magnetic field when it is tilted
along the $x$ or $y$ directions, i. e. for $\varphi=\pi/2$ ($B_x=0$)
and for $\varphi=0$ ($B_y=0$). (b) The magnetization recalculated with
the Rashba SOC parameter increased by 50\%. In both cases the Dresselhaus 
SOC strength is $\beta=3$ meV nm}
\vspace{8mm}
\label{fig:Magnetization}
\end{figure}

For a homogeneous 2DES, i. e. in the absence of the periodic modulation,
the magnetization has saw-teeth-like de Haas - van Alphen oscillations
vs. the $z$ component of the magnetic field, corresponding to the
motion of the Fermi energy within a degenerated Landau level, followed
by an abrupt jump to the adjacent level. For the modulated system,
the magnetization has a positive or a negative peak when the Fermi
level touches the bottom or the top of a Landau band, respectively
\cite{Gudmundsson2000}.  

In Fig.\ \ref{fig:Magnetization} we show the magnetization calculated
for the two present situations of interest: when the magnetic field
is tilted in the $x$ or in the $y$ direction, i. e.  when $\varphi=0$
and $\varphi=\pi/2$, respectively. The peaks indicate the van Hove
singularities of the Landau bands. Comparing to the resistivities,
Fig.\ \ref{fig:Resistivity}, our calculations predict that the van
Hove singularities can be much better resolved in magnetization
measurements, even if the Landau bands are partially overlapped, as
in Fig.\ \ref{fig:Landau_bands}, which is the situation in all our
calculations.

The oscillations of the magnetization in the modulated 2DES are also quite
complex, being determined by the band structure, but indirectly influenced
by the Coulomb many-body effects. The exchange interaction may lead to larger
magnetization fluctuations than expected for noninteracting electrons
\cite{Gudmundsson2000,Meinel2001}.  Consistently with the dispersion
of the Landau bands, due to the presence of the SOC, the magnetization also
depends on the azimuthal angle $\varphi$ of the in-plane magnetic field.
\red{Like in the transport calculations we considered two values of the
Rashba SOC strength for comparison.  Since the van Hove singularities
of the Landau bands may shift either up or down with increasing $\alpha$,
the magnetization peaks may shift either up or down on the magnetic field scale,
irrespective of the orientation of the magnetic field.}

\section{Conclusions}

We calculated the Landau bands produced in a 2DES by a periodic
unidirectional potential in the presence of a tilted magnetic field,
with a strong component perpendicular to the 2DES.  In the absence of
SOC, when the spin contributes only with the Zeeman coupling to the total
magnetic field, the Landau bands do not depend on the orientation of the
in-plane component of the magnetic field relatively to the direction of
the modulation. They depend on the polar angle $\theta$, but not on the
azimuthal angle angle $\varphi$, indicated in Fig. \ref{fig:System}.

The interplay of the SOC, the unidirectional periodic modulation, and
the in-plane magnetic field, result in Landau bands whose dispersion
explicitly depends on the projections of the magnetic field on axes
perpendicular and parallel to the modulation, respectively. Therefore
measured observables like electrical resistances or magnetization, in the
quantum Hall regime, in a tilted magnetic field,  should be sensitive
to the rotation of the field, conically, around the normal to the 2DES
plane, i. e. to the azimuthal angle angle $\varphi$. \red{Such experimental 
investigations are reported for the homogeneous 2DES \cite{wilde13}, but,
to the best of our knowledge, not for the modulated 2DES. We are also not aware of 
alternative experimental evidence of the asymmetric energy dispersion due
to the interplay of SOC and an in-plane magnetic field.}

The results that we have shown for the electrical resistance and magnetization 
are only qualitative. As long as the SOC effects are weak, 
they become significant only for Landau bands with strong dispersion,
produced by a modulation potential considerably larger than the cyclotron
energy.  Given the complexity of the Landau bands, especially due to
the inclusion of the Coulomb effects, it is difficult to make reliable
quantitative predictions on the anisotropy of the magnetization or
resistances (which are complex observables themselves) in order to relate
some features with the SOC strength.  
\red{The asymmetry under magnetic field orientation could become a diagnostic
tool for an estimate of the SOC presence and intensity in the 2DES.
Since the van Hove singularities at the band edges can be well resolved in 
the magnetization we expect that the SOC effects on the Landau bands could 
be better observable in magnetization, rather than in magnetotransport measurements.}

Finally, the observation of this SOC effect in the magnetoresistance
and magnetization experiments proposed in this paper, is likely to depend
not only on the SOC strength itself, but also on the sample characteristics,
such as disorder or electron concentration, details that can only be established 
in future experimental work.  \\

\end{document}